\documentclass[12pt]{article}
\usepackage{graphicx}
\usepackage[utf8x]{inputenc}
\usepackage{units}
\usepackage{amssymb}

\newcommand{\Vud}{$V_\mathrm{ud}$\ }

\newcommand\pubnumber{} 
\newcommand\pubdate{\today}

\def\ati{Atominstitut, Technische Universität Wien, Stadionallee 2, 1020 Wien, AUSTRIA}
\def\heidelberg{Physikalisches Institut\\
Universität Heidelberg, Im Neuenheimer Feld 226, D-69120 Heidelberg, GERMANY}
\def\supp{\footnote[1]{Work supported by the Priority Programme SPP~1491 of
the German Research Foundation~(DFG) under contract MA4944/1.}}
\def\suppo{\footnote[2]{Work supported by the Priority Programme SPP~1491 of the German Research Foundation~(DFG) and by the
Austrian FWF under contract I529-N20.}}

\def\Title#1{\begin{center} {\Large #1 } \end{center}}
\def\Author#1{\begin{center}{ \sc #1} \end{center}}
\def\Address#1{\begin{center}{ \it #1} \end{center}}

\newcommand\pubblock{\rightline{\begin{tabular}{l} \pubnumber\\
         \pubdate  \end{tabular}}}
\newenvironment{Abstract}{\begin{quotation}  }{\end{quotation}}
\newenvironment{Presented}{\begin{quotation} \begin{center}
             PRESENTED AT\end{center}\bigskip
      \begin{center}\begin{large}}{\end{large}\end{center} \end{quotation}}
\def\Acknowledgements{\bigskip  \bigskip \begin{center} \begin{large}
             \bf ACKNOWLEDGEMENTS \end{large}\end{center}}

\def\beq{\begin{equation}}
\def\eeq#1{\label{#1}\end{equation}}
\def\eeqn{\end{equation}}

\def\beqa{\begin{eqnarray}}
\def\eeqa#1{\label{#1}\end{eqnarray}}
\def\eeqan{\end{eqnarray}}

\let\bar=\overbar

\def\M{{\cal M}}
\def\O{{\cal O}}

\def\Dslash{\not{\hbox{\kern-4pt $D$}}}
\def\dslash{\not{\hbox{\kern-2pt $\del$}}}

\def\msb{{\bar{\ssstyle M \kern -1pt S}}}

\begin{document}
\begin{titlepage}
\pubblock

\vfill
\Title{Measurement of the Axial-Vector Coupling Constant $g_A$ in Neutron Beta Decay}
\vfill

\Author{Bastian Märkisch\supp}
\Address{\heidelberg}
\Author{Hartmut Abele\suppo}
\Address{\ati}
\vfill
\begin{Abstract}
The matrix element \Vud of the CKM matrix can be determined by two independent measurements in neutron decay: the neutron lifetime $\tau_n$ and the ratio of coupling constants $\lambda=g_A/g_V$, which is most precisely determined by measurements of the beta asymmetry angular correlation coefficient~$A$. We present recent progress on the determination of these coupling constants.
\end{Abstract}
\vfill
\begin{Presented}
8th International Workshop on the CKM Unitarity Triangle (CKM 2014), \\ Vienna, Austria, \\ September 8-12, 2014
\end{Presented}
\vfill
\end{titlepage}
\def\thefootnote{\fnsymbol{footnote}}
\setcounter{footnote}{0}

\section{Introduction}

To obtain the matrix element \Vud from the decay of the free neutron only two separate inputs are required. These are the neutron lifetime $\tau_n$ and the ratio of axial vector and vector coupling constants $\lambda=g_A/g_V$, which can be determined by measurements of angular correlations in neutron decay \cite{Jackson57,Wilkinson82}. \Vud can then be determined by \cite{Marciano06}
\begin{equation}
\label{eq:Vud}
\left|V_\mathrm{ud}\right|^2 = \frac{(4908.7 \pm 1.9)\texttt{s}}{\tau_n \left(1+3\lambda^2 \right)},
\end{equation}
where the numerator includes all constants, with the Fermi coupling constant precisely determined in muon decay, and the theoretical uncertainty of the radiative corrections (see also Ref.~\cite{Czarnecki04}).

Assuming vector current conservation, the axial vector coupling constant can be determined within the Standard Model from a variety of angular correlation measurements \cite{Jackson57, Wilkinson82}. The most precise determination comes from measurements of the beta asymmetry $A$ correlation coefficient, which describes the correlation between neutron spin and electron momentum. To leading order $A_0$ this asymmetry is given by
\begin{equation}
A_0 = \frac{-2 \left( \lambda^2 - |\lambda| \right)}{1 + 3 \lambda^2}.
\end{equation}
Due to the equally high sensitivity on $\lambda$, the determination of the electron-neutrino angular correlation $a$ is another candidate. This work is an update of earlier presentations on the unitarity triangle~\cite{Abele03A,Blucher05,Maerkisch11,Zimmer13}, the unitarity of the CKM matrix~\cite{Abele03}, and other reviews~\cite{Severijns06,Abele08,Dubbers12,Abele00}.

\section{Recent Results}

The most recent determinations of the beta asymmetry come from the UCNA Collaboration and the \textsc{Perkeo~II} Collaboration. The decay of polarized neutrons in a strong magnetic field is analysed by electron spectroscopy with a solid angle coverage of $2 \times 2\pi$. In these experiments backscattering of electrons from the detectors~\cite{Martin03,Martin06}, a serious source of error in $\beta$-spectroscopy, is strongly suppressed by a decrease in magnetic field strength towards the detectors and detection of backscattered electrons in the second $2\pi$ detector~\cite{Abele93,Schumann08b,Dubbers14}. A backscatter suppression spectrometer is also described in~\cite{Wiet05}.

The UCNA Collaboration uses polarized ultracold neutrons at the Los Alamos Neutron Science Center (LANSCE).  Improvements in the experiment have led to reductions in both statistical and systematic uncertainties.  UCNs were polarized by a $6\,\unit{T}$ prepolarizer magnet and a $7\,\unit{T}$ primary polarizer.  The spin state is controlled by an adiabatic fast passage spin flipper.  Upstream of the prepolarizer magnet, a gate valve separates the UCN source from the experimental apparatus. Polarized UCNs enter the superconducting spectrometer and are confined in a $3\,\unit{m}$ long, $12.4\,\unit{cm}$ diameter diamond-like carbon coated copper tube (decay trap).  Decay electrons spiral towards one of two identical electron detectors along a $0.96\,\unit{T}$ magnetic field oriented parallel to the decay trap.  Each detector consists of a $16 \times 16 \unit{cm}^{2}$  low-pressure multiwire proportional chamber backed by a $15\,\unit{cm}$ diameter plastic scintillator with photomultiplier readout~\cite{Mendenhall13}.
The most recent result \cite{Mendenhall13} is
\begin{equation}
A_0 = -0.11954(55)_\mathrm{stat}(98)_\mathrm{syst},
\end{equation}
corresponding to
\begin{equation}
\lambda = g_A/g_V  = -1.2756(30)\label{eq:lambdaU}.
\end{equation}
Earlier publications can be found in Refs.~\cite{Pattie09,Liu10,Plaster12}.

The \textsc{Perkeo~II} Collaboration \cite{Abele02,Abele08} uses a strong cold neutron beam at the ballistic supermirror guide H113~\cite{Abele06} at Institut Laue-Langevin (ILL). An X-arrangement of two supermirror polarizers~\cite{Kreuz05} makes it possible to achieve an unprecedented degree of neutron polarization $P = 99.7(1)\%$ over the full cross-section of the beam. The main component of \textsc{Perkeo~II} is a superconducting split  pair
configuration with a coil diameter of about $1\,\unit{m}$, producing a $1\,\unit{T}$ field oriented perpendicular to the neutron beam.  Neutrons pass through the spectrometer, whereas decay electrons and protons are guided by the magnetic field to either one of two detectors. A description of the detector system can be found in~\cite{Abele09}. The electron detectors are scintillation detectors with photomultiplier readout. The concept of loss free electron spectroscopy is described in Ref.~\cite{Abele93}.

The combined results of the new~\cite{Mund13} and previous \cite{Abele97,Abele02} \textsc{Perkeo~II} experiments are
\begin{equation}
A_0 = -0.11951(50)
\end{equation}
and
\begin{equation}
\lambda = -1.2748(13).\label{eq:lambdaP}
\end{equation}
In the above averages correlations of systematic errors in the experiments are accounted for. Conservatively, errors concerning detector calibration and uniformity, background determination, edge effect, and the radiative correction were considered correlated on the level of the smallest error of all three experiments.

Other experiments \cite{Yero97,Liaud97,Bopp86} gave significantly lower values for $|\lambda|$ compared to Eq.~\ref{eq:lambdaU} and Eq.~\ref{eq:lambdaP}. However, in these experiments large corrections had to be applied for neutron polarization, magnetic mirror effects, solid angle, or background, which were in the $15\%$ to $30\%$ range. A history plot including corrections is found in Ref.~\cite{Abele08}.

Complementary to electron asymmetry measurements, several projects (\emph{a}SPECT~\cite{Baessler08,Konrad09}, aCORN~\cite{Wiet09}, Nab~\cite{Nab}, PERC~\cite{Dubbers08,Konrad12}) aim to derive the ratio of coupling constants $\lambda$ with competitive precision from the electron-neutrino correlation coefficient $a$.

For a determination of coefficients $B$~\cite{Kreuz05,Schumann07} and $C$~\cite{Schumann08}, a combined electron and proton detector has been installed in each hemisphere~\cite{Abele09}. So far measurements  of the coefficient $C$ are not competitive in determining $\lambda$, yet.

We combine the results of \cite{Yero97,Liaud97,Bopp86} and ~\cite{Mostovoi01,Schumann08} and the new results of \cite{Mendenhall13,Mund13} to obtain
\begin{equation}
\label{eq:ourlambda}
\lambda = -1.2723(21),\qquad S=2.04,
\end{equation}
were we have scaled the error bar according to standard PDG procedures.

As the theoretical considerations are concerned, neutron $\beta$-decay is understood on the $10^{-4}$ level or better; the analysis of lifetime, energy spectra, and angular distributions of the neutron beta--decay was carried out within the standard quantum field theory at the rest frame of the neutron and in the non--relativistic approximation for the proton by taking into account the radiative corrections to order $\alpha$, the proton recoil corrections, and the ``weak magnetism'' to order $1/M$, where $M = (m_n + m_p)/2$ is the average mass~\cite{Ivan13a,Ivan13b,Ivan13c}.
\section{Neutron Lifetime}

At this workshop the status of the neutron lifetime has been presented by~\cite{Wietfeldt}. Here, we determine the neutron lifetime from $\lambda$ and $\overline{\mathcal{F}t}$ value from nuclear beta decay. Assuming the $V-A$ structure of the Standard Model,
\begin{equation}
\tau_\mathrm{n} = \frac{2}{\ln 2} \frac{\overline{\mathcal{F}t}}{f_R\,(1+3\lambda^2)},
\end{equation}
where the phase space factor $f_R = 1.71385(34)$ \cite{Konrad10} includes radiative corrections. We use the average of the most resent results of Eq.~\ref{eq:lambdaU} and Eq.~\ref{eq:lambdaP} and $\overline{\mathcal{F}t} = 3071.81(83)$ \cite{Hardy09} to derive the neutron lifetime
\begin{equation}
\tau_\mathrm{n} = 880.1(1.4)\,\unit{s}.
\label{eq:ourlifetime}
\end{equation}
This result is in agreement with and nearly as precise as the current world average $\tau_{\mathrm{n}} = 880.3(1.1)\,\unit{s}$ \cite{PDG2014}, which includes a scale factor of the error bar of $1.9$.

\section{Summary and Outlook}

The angular correlation measurements are pursued by a lively community. Currently, the  \textsc{Perkeo~III}, UCNA, UCNB, \textit{a}{SPECT} and aCORN experiments are running or analysing data.

Ref.~\cite{Baessler14} describes the beta decay program at the Spallation Neutron Source at Oak Ridge National Laboratory, and puts it into the context of other (planned) measurements of neutron beta decay observables.

A program of precision studies of neutron beta decay is planned with the Nab spectrometer. One aim is to perform a precise measurement of $a$, the electron-neutrino correlation parameter, and $b$, the Fierz interference term in neutron beta decay, at the Fundamental Neutron Physics Beamline at the Spallation Neutron Source, using a novel electric/magnetic field spectrometer and detector design. The experiment is aiming at the $10^{-3}$ accuracy level on $a$, and will provide an independent measurement of $\lambda$~\cite{Nab}.

The \textsc{Perkeo III} spectrometer \cite{Maerkisch09,Maerkisch14} increases the size of the active volume by almost two orders of magnitude, which makes it feasible to use a pulsed neutron beam. Beam related background can now be measured under the same conditions as the decay signal itself and can thus be fully subtracted.

The instrument PERC~\cite{Dubbers08,Konrad12}, which is under development by an international collaboration, follows a radically novel concept to improve systematics and statistics: to maximise the phase space of the neutrons, a neutron guide consisting of non-depolarising mirrors in a strong magnetic field is used as \emph{active volume}.  A magnetic filter is then used to limit the phase space of the emerging electrons and protons~\cite{Wang12}.  A detailed analysis shows that all relevant sources of systematic error can be controlled on the $10^{-4}$ level or better, an improvement by one order of magnitude in comparison to existing spectrometers.

\Acknowledgements
The authors would like to thank the organizers and conveners for their kind invitation to this inspiring conference.

\end{document}